\documentclass{aa}

\usepackage{graphicx}

\usepackage{txfonts}

\usepackage{xcolor}

\usepackage{soul}

\begin{document}

   \title{Separating source-intrinsic and Lorentz invariance violation induced delays in the very high-energy emission of blazar flares}

   \author{C. Levy\inst{1,2}
          \and
         H. Sol\inst{2}
         \and
         J. Bolmont\inst{1}
         }

   \institute{Sorbonne Universit\'e, CNRS/IN2P3,\\ Laboratoire de Physique Nucl\'eaire et de Hautes Energies, LPNHE,\\ 4 Place Jussieu, F-75005 Paris, France\\
        \email{bolmont@lpnhe.in2p3.fr}
        \and
        LUTH, Observatoire de Paris, PSL Research University,\\ CNRS, Universit\'e Paris Cit\'e,\\ 5 Place Jules Janssen,
F-92190 Meudon, France\\
\email{helene.sol@obspm.fr}
             }

   \date{Received XXX; accepted YYY}

\abstract{}{
The aim of the present study is to explore how to disentangle energy-dependent time delays due to a possible Lorentz invariance violation (LIV) at Planck scale from intrinsic delays expected in standard blazar flares.} {
We first characterised the intrinsic time delays in BL Lacs and flat-spectrum radio quasars in standard one-zone time-dependent synchrotron self-Compton or external Compton models, during flares produced by particle acceleration and cooling processes. We simulated families of flares with both intrinsic and external LIV-induced  energy-dependent delays. Discrimination between intrinsic and LIV delays is then investigated in two different ways. A technique based on Euclidean distance calculation between delays obtained in the synchrotron and in the inverse-Compton spectral bumps is used to assess their degree of correlation. A complementary study is performed using spectral hardness versus intensity diagrams in both energy ranges.
} {
We show that the presence of non-negligible LIV effects, which essentially act only at very high energy (VHE), can drastically reduce the strong correlation expected between the X-ray and the VHE gamma-ray emission in leptonic scenarios. The LIV phenomenon can then be hinted at measuring the Euclidean distance $d_{E}$ from simultaneous X-ray and gamma-ray flare monitoring. Large values of minimum distance $d_{E,min}$ would directly indicate the influence of non-intrinsic time delays possibly due to LIV in SSC flares. LIV effects can also significantly modify the VHE hysteresis patterns in hardness-intensity diagrams and even change their direction of rotation  compared to the X-ray behaviour. Both observables could be used to discriminate between LIV and intrinsic delays, provided high-quality flare observations are available.}
{}


\titlerunning{Separating source-intrinsic and LIV induced delays in the VHE emission of blazar flares}

   \maketitle
%

\section{Introduction}

One of the most challenging ambitions of modern physics is to unify the realms described by quantum field theory and general relativity. This unification could be achieved through a still-in-the-making unique theory of quantum gravity (QG). %
The main problem encountered in the process of developing this new fundamental theory is the lack of experimental input, due to the fact that the energy scale involved in quantum gravity theories is believed to be extremely high, of the order of the Planck energy, $E_P \sim 10^{19}$ GeV. Moreover, extracting observable predictions from still tentative models has been notoriously difficult. 
For a more detailed discussion about this topic, as well as a detailed review of the state of the art of the different theoretical and experimental efforts currently ongoing in QG phenomenology, we refer to \cite{COST}.

Nevertheless, a Lorentz invariance violation (LIV) or deformation appears as a possible effect that could be tested with present-day gamma-ray experiments. These effects are usually represented by a modified dispersion relation for photons in vacuum of the form
\begin{equation}
E^2 \simeq p^2c^2 \times \left[ 1 \pm \sum_{n=1}^{\infty} \left( \frac{E}{E_{QG, n}} \right)^n \right],
\label{eq:disprel}
\end{equation}
where $E$ and $p$ are the photon energy and momentum, $c$ is the standard (low-energy) speed of light, $n$ is the correction order, and $E_{QG, n}$ is the  quantum gravity energy scale, which may be different for different values of $n$. The correction terms are expected to become significant for $E \sim E_{QG, n} \sim E_P$, but should be present at lower energies. The $\pm$ sign accounts for both subluminal and superluminal effects. The modified dispersion relation (\ref{eq:disprel}) leads to energy-dependent velocities inducing delays $\Delta t$ in the arrival time of gamma-ray photons with different energies, assuming they are emitted at the same time and location:
\begin{equation}
\Delta t = \pm \frac{n+1}{2} \frac{\Delta E^n}{E_{QG,n}^n} \times \kappa_n(z).
\end{equation}
Here $\Delta E^n = | E^n_1 - E^n_2 |$ is the energy difference at order $n$, $z$ the redshift, and $\kappa_n(z)$ an increasing function of redshift accounting for cosmological effects. Variation with cosmic distance then appears as a distinctive feature of QG-induced delays. Instead of working with $\Delta t$, a common practice is to use the time delay over energy difference, denoted $\tau_n$, and defined as
\begin{equation}
\label{eq:tau}
\tau_n = \frac{\Delta t}{\Delta E^n} = \pm \frac{n+1}{2\,H_0\,E_{QG,n}^n} \times \kappa_n(z),
\end{equation}
where $H_0$ is the Hubble constant. In the following we   deal only with this quantity, but for completeness, we note  that $\kappa_n$ has been mostly taken from \cite{Jacob2008} so far, where the expression 
\begin{equation}
\label{eq:kappa}
\kappa_n(z) = \int_0^z \frac{(1+z')^n}{\sqrt{\Omega_m\,(1+z')^3 + \Omega_\Lambda}}\ dz'
\end{equation}
is obtained, with $\Omega_m$ and $\Omega_\Lambda$ respectively the matter density and the dark energy density of the Universe.
In the context of gamma-ray astronomy, the quantity $\tau_n$ is usually given in units of s\,TeV$^{-n}$ or s\,GeV$^{-n}$. Provided very high-energy photons propagate through very long distances, such delays could be detectable with present-day experiments, at least for a linear correction to the dispersion relation ($n = 1$). We   only consider this case in the following, noting $\tau \equiv \tau_{n = 1}$, expressed in s\,TeV$^{-1}$.

It should be noted that the above description sticks to simple cases when considering LIV delays. In particular, Eq.~\ref{eq:kappa} is only one possible way to account for distance in LIV delay computation \citep[see e.g.][]{Rosati2015, Pfeifer2018}
and other possible QG-related LIV effects, such as the modification of extra-galactic background light (EBL) absorption \citep[see][§5.3]{COST}, are neglected. Moreover, classical intrinsic effects arising in the source can limit the possibility of having bunches of gamma-ray photons emitted from the same location and at the same time, which is the primary focus of the present paper.

Following Eq.~\ref{eq:tau}, a strategy currently in use to search for LIV signatures is to look for energy-dependent time delays in the gamma-ray signal coming from remote and variable cosmic sources, such as flaring active galactic nuclei (AGN), pulsars, and gamma-ray bursts \citep[GRBs,][]{COST,terzic2021}. These sources are detected at MeV--GeV energies by satellites and at TeV energies with ground-based gamma-ray instruments, such as imaging atmospheric Cherenkov telescopes (IACTs). 
However, measured delays can include time lags that can be generated by classical radiative processes at the sources, or any other effect that could modify photon propagation depending on its energy. The latter effects can be due, for example (but not only), to exotic plasma effects at the source \citep{Latorre1995,Myers2003,Albert2008} or to the delayed production of secondary photons in cascades deflected by extragalactic magnetic fields \citep{Taylor2011}. In the present paper, we focus on the former,  called the  source intrinsic effects, specifically dealing with the case of blazar flares. Such classical intrinsic effects can lead to energy-dependent time delays, which have been shown to be non-negligible at very high energy (VHE) in standard flare models \citep{Perennes} for a large domain of initial parameters, and would need to be discriminated from LIV-induced delays in the case of a significant lag detection. 

To date, no significant lag has been measured or limits on $E_{QG}$   derived, except on one occasion for a flare of the blazar \object{Mkn~501} recorded by MAGIC in July 2005 \citep{4min}. It is probable, however, that a new generation of instruments with improved sensitivity will lead to more frequent lag detection. At this point, it will be necessary to disentangle between source intrinsic effects and LIV-induced delays.

The first way to achieve that goal relies on the fact that source intrinsic time delays are not expected to directly depend on the distance of the source. On the contrary, as described by Eq.~\ref{eq:tau}, the LIV time delay has an explicit dependence on the redshift, expressed by the function $\kappa_n(z),$ which continuously increases with $z$. So it is expected that detected delays should significantly vary with $z$ if they are due to LIV, while they have no immediate reasons to do so if they are due to intrinsic effects. It is the variability timescale of the flares (typically from minutes to hours) that allows intrinsic delay detection, and determines (by the causality argument) the physical size of the emitting zone. This size directly governs, in turn, the particle acceleration and cooling timescales and intrinsic time delays. So in a typical sample of blazar flares, the observed intrinsic delays are not supposed to depend primarily on distance.\footnote{At second order, the possibility of a complex Malmquist bias, taking into account Doppler boosting effects, would deserve a further analysis, beyond the scope of the present study.} Therefore, a discrimination strategy relies on population studies involving a large number of sources over a broad range of redshifts. Similar analyses have already been published using multiple GRBs observed by satellites \citep[see e.g.][and references therein]{Ellis2019}, leading to lower limits on $E_{QG}$ up to $10^{17}$~GeV for $n = 1$, and an effort is currently ongoing to perform a population study with all sources (AGN, GRBs, and pulsars) detected by present-day IACT arrays \citep{HMV}. 

Another complementary way to separate intrinsic and LIV-induced delays relies on the study of the intrinsic effects through the modelling of source emission mechanisms. A first attempt in that direction was presented by \cite{Perennes} for flares of BL Lac sources. 
Starting from a simple one-zone purely leptonic synchrotron self-Compton (SSC) scenario, and using conservative physical parameters to describe the emitting plasma, the authors showed how energy-dependent intrinsic lags are produced during single flares generated by acceleration and cooling processes. They identified two main regimes, depending on the domain of parameters, one where the lags increase with energy at VHE (with lower energies leading to higher ones), and one where the lags decrease with energy at VHE (with higher energies leading to lower ones). In the first regime (called acceleration-driven), flares are globally dominated by a relatively slow acceleration mechanism and reach their peak before electrons reach their maximum energy, while in the second regime (called cooling-driven), they are globally dominated by the cooling effect and reach their peak after electrons reach their maximum energy. 
These intrinsic lags can be large at VHE, reaching several hundreds of seconds per TeV, meaning that present-day observations could already be used to constrain this kind of model, at least for very bright flares. It is also found that intrinsic lags vary with time during the flare. Since LIV delays obviously do not, this may offer another way to separate LIV and intrinsic delays. However, measuring the evolution of the delay in VHE during a single BL Lac flare seems completely out of reach at the moment, and may even be challenging with the next generation of instruments such as the Cherenkov Telescope Array\,\footnote{\url{https://www.ctao.org}} (CTA).

In this paper we explore new ways to possibly disentangle potential LIV lags from intrinsic delays. The results shown were obtained with a software library called the AGN Evolution Simulator (AGNES),\,\footnote{The code can be downloaded from \url{https://gitlab.in2p3.fr/julien.bolmont/AGNES}.} specifically developed for that purpose by C.~Levy. In Sect.~\ref{sec:intdelays} we recall and further develop the study by \cite{Perennes}, examining the impact on intrinsic delays of adiabatic expansion and external inverse-Compton emission phenomena expected in flat-spectrum radio quasars (FSRQs). 
In Sect.~\ref{sec:addingliv} LIV effects are added to the emission model output, allowing the simultaneous description and direct comparison of the two types of delays. We show how multi-wavelength monitoring of blazar flares, covering X-ray and gamma-ray domains, could offer direct possibilities to distinguish the presence of an additional lag potentially due to LIV over classical intrinsic delays, even in the case where only one source is considered at one particular redshift. Two techniques are proposed to combine simultaneous X-ray and gamma-ray data with the goal to discriminate LIV from intrinsic effects. These techniques are tested on the model output. Conclusions and perspectives are given in Sect.~\ref{sec:conc}.

\section{Characterisation of intrinsic time delays at very high energy in blazars} \label{sec:intdelays}

Several time dependent non-thermal emission models for radio, X-ray, and gamma-ray flares of AGN proposed in the literature lead to the existence of intrinsic time delays generated from radiation mechanisms \citep[e.g.][]{Model1,Model2,Model3,Model4,Katarzynski2003,Lewis}. 
These types of leptonic models are extensively used in the literature to describe the temporal evolution of flares generated by particle injection, acceleration, and cooling. They are fairly constrained and are considered reference models by the community since they can reproduce simple blazar flares reasonably well, at least for currently available datasets. For VHE flares varying quickly on timescale $t_0$, the size of the emission zone is limited to $R \leq c t_0 \delta_b/(1+z)$ by the causality argument, where $\delta_b$ is the Doppler boost of the blob, and emission models dominated by one compact zone are often preferred. We therefore adopt here a leptonic scenario to further analyse the intrinsic time delays at high and very high energies, describing the VHE emission zone as a single optically thin small plasma blob travelling along the blazar's jet with a bulk Lorentz factor $\Gamma$, and harbouring an electron-positron population submitted to acceleration and cooling processes. 

\subsection{Leptonic flare scenario} 

In the reference frame of the emission zone, the evolution of the electron-positron spectrum can be described by the following differential equation: 
\begin{equation}
\label{eq:ED_adia}
\frac{\partial N^*(t,\gamma)}{\partial t} = \frac{\partial}{\partial \gamma} \left \{ \left[ C_{\mathrm{rad}}(t)\gamma^2 - (C_{\mathrm{acc}}(t) - C_{\mathrm{adia}}(t))\gamma \right] N^*(t,\gamma) \right \}.
\end{equation}Here $t$ and $\gamma$ are the time and the individual particle Lorentz factor; $C_{\mathrm{acc}}(t) \propto A(t)$
is the acceleration term, with $A(t)$ a function defined as
\begin{equation}
A(t) = A_0 \left( \frac{t_0}{t} \right)^{m_a},
\end{equation}
with $A_0$ the initial acceleration amplitude and $m_a$ the evolution index. The radiative term $C_{\mathrm{rad}}(t)$ is defined as $C_{\mathrm{rad}}(t) \equiv C^\mathrm{SSC}_{\mathrm{rad}}(t) + C^{ext}_{\mathrm{rad}}$, 
where $C^{ext}_{\mathrm{rad}}$ stands for the particle radiation and cooling due to inverse-Compton losses on any external photon field, here assumed to be constant. The other term $C^\mathrm{SSC}_{\mathrm{rad}}(t) \propto B(t)^2$ accounts for the particle radiation and cooling due to the synchrotron losses and the synchrotron-self-Compton losses, here assumed smaller than and proportional to synchrotron losses \citep{Katarzynski2001}. The evolution of the magnetic field $B(t)$ is parametrised as 
\begin{equation}
B(t) = B_0 \left( \frac{t_0}{t} \right)^{m_b},
\end{equation}
with $B_0$ the initial magnetic field strength and $m_b$ the evolution index. The parameter $t_0$ is the characteristic evolution time, taken as the time needed for a sound wave to propagate through the blob of relativistic plasma of radius $R$. The quantity $C_{\mathrm{adia}}(t)$ in Eq.~\ref{eq:ED_adia} describes the effect of slow adiabatic expansion (or compression) of the compact spherical emission zone with radius $R(t)$ parametrised as
\begin{equation}
R(t) = R_0 \left( \frac{t_0}{t} \right)^{-m_r},
\end{equation}
with $R_0$ the blob's initial radius, and $m_r$ the evolution index. The quantity $N^*(t,\gamma)$ is the number density per unit Lorentz factor $\gamma$, modified to take into account the blob's radius evolution index, which reads
\begin{equation}
\label{eq:newN}
N^*(t,\gamma) = N(t,\gamma)\left( \frac{t_0}{t} \right)^{3m_r},
\end{equation}
where $N(t,\gamma)$ is the particle number density. The sign attributed to the evolution index $m_r$ determines whether it contributes to expanding or compressing the blob's volume. If positive, the blob is expanding and a loss of energy occurs, working against the acceleration; if negative, the blob is compressing and contributes to the energy gains complementing the acceleration processes.

  \begin{figure*}
  \centering
  \includegraphics[width=0.97\linewidth]{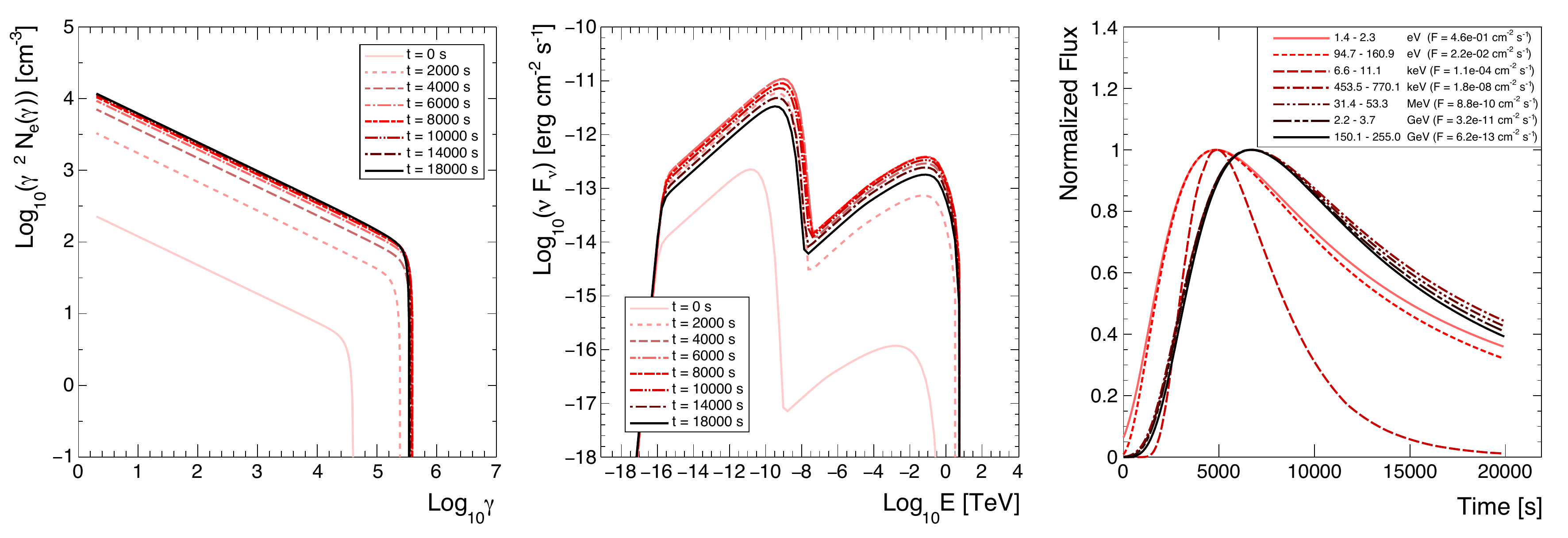}
  \caption{Distributions obtained from solving the differential equation governing the lepton number density, for the set of reference input parameters given in Table~\ref{tab:parameters}. No stationary state contribution is included here. Left: Lepton spectra temporal evolution. Centre: Spectral energy distributions of temporal evolution, generated as seen by the observer. Right: Normalised multi-wavelength light curves.}\label{fig:spectra}
    \end{figure*}

In BL Lacs, SSC radiation is expected to dominate in the absence of a bright accretion disc, a broad line region (BLR), and a dust torus. Introducing adiabatic processes and external inverse-Compton (EIC) emission allows  further exploration beyond the BL Lac case analysed by \cite{Perennes}, especially extending it to FSRQs, where the EIC emission should play a significant role. The additional number of photons brought by the external radiation fields contribute to generate VHE photons and boost the flux in the SED inverse-Compton component. External photon fields generated by a bright disc, a BLR, and a torus in FSRQ are expected to dominate the EIC component respectively one after the other with the increase in the distance $d$ of the emitting blob relative to the central black hole, and can be roughly modelled with a blackbody emission centred on a specific temperature. To illustrate their effect on intrinsic time delays, we consider here a static photon field, generated by the accretion disc and reprocessed by the clouds of the BLR located at the distance $d$ with typically $0.01\ \mathrm{pc} < d < 0.1\ \mathrm{pc}$. This results in an ambient radiation field that is Doppler boosted as seen in the blob's frame when moving towards the BLR clouds. The cooling coefficient $C^{\mathrm{ext}}_{\mathrm{rad}}$ is taken to be proportional to the energy density $U_{\mathrm{ext}}$, as described in \cite{Takahara}
\begin{equation}
\label{eq:EIC}
C^{\mathrm{ext}}_{\mathrm{rad}} \propto U_{\mathrm{ext}} = \Gamma_b^2 \frac{\tau L_{AD}}{4 \pi d^2 c},
\end{equation}
with $\Gamma_b$ the Lorentz factor of the plasma blob, $L_{AD}$ the accretion disc luminosity, $d$ the distance between the accretion disc and the plasma blob, and $\tau$ the reprocessing efficiency combined with the covering factor of the BLR. A significant fraction of the external photons (those that are directed or reflected head-on towards the relativistic plasma blob) are strongly blueshifted in the source's rest frame. 
In particular, photons reprocessed by the BLR can reach keV energies (X-rays) and therefore be boosted up to TeV energies through inverse-Compton interactions with the blob's leptons. For illustrative purposes, we adopt hereafter a typical value for the reprocessed luminosity of $\tau L_{AD} = 10^{41}$ erg\,s$^{-1}$, and only vary the distance $d$ from 0.03 pc to infinity, to analyse stronger and weaker EIC effects. We note that in reality at large distances (typically $d > 0.1$ pc), once the blob has crossed the BLR, the ambient radiation due to the dust torus is supposed to dominate the EIC component. Here we explore only the qualitative effect of EIC emissions on the temporal time delays, without aiming to reproduce in detail a specific AGN.

\subsection{Intrinsic time delays in the gamma-ray range}

Solving Eq.~\ref{eq:ED_adia} gives access to the evolution of lepton spectra with time, to the evolution of spectral energy distributions, and to the associated multi-wavelength light curves, after transformation from the source frame to the observer frame. Examples of these distributions can be found in Fig.~\ref{fig:spectra} for a typical weak flare based on a pure one-zone SSC emission scenario, which can serve as a good proxy for stationary states and VHE flares of BL Lacs. A standard extragalactic background light (EBL) model provided by \cite{Kneiske2002,Kneiske2004} is taken into account for all the work presented in this paper.

  \begin{figure}
  \centering
  \includegraphics[width=0.9\linewidth]{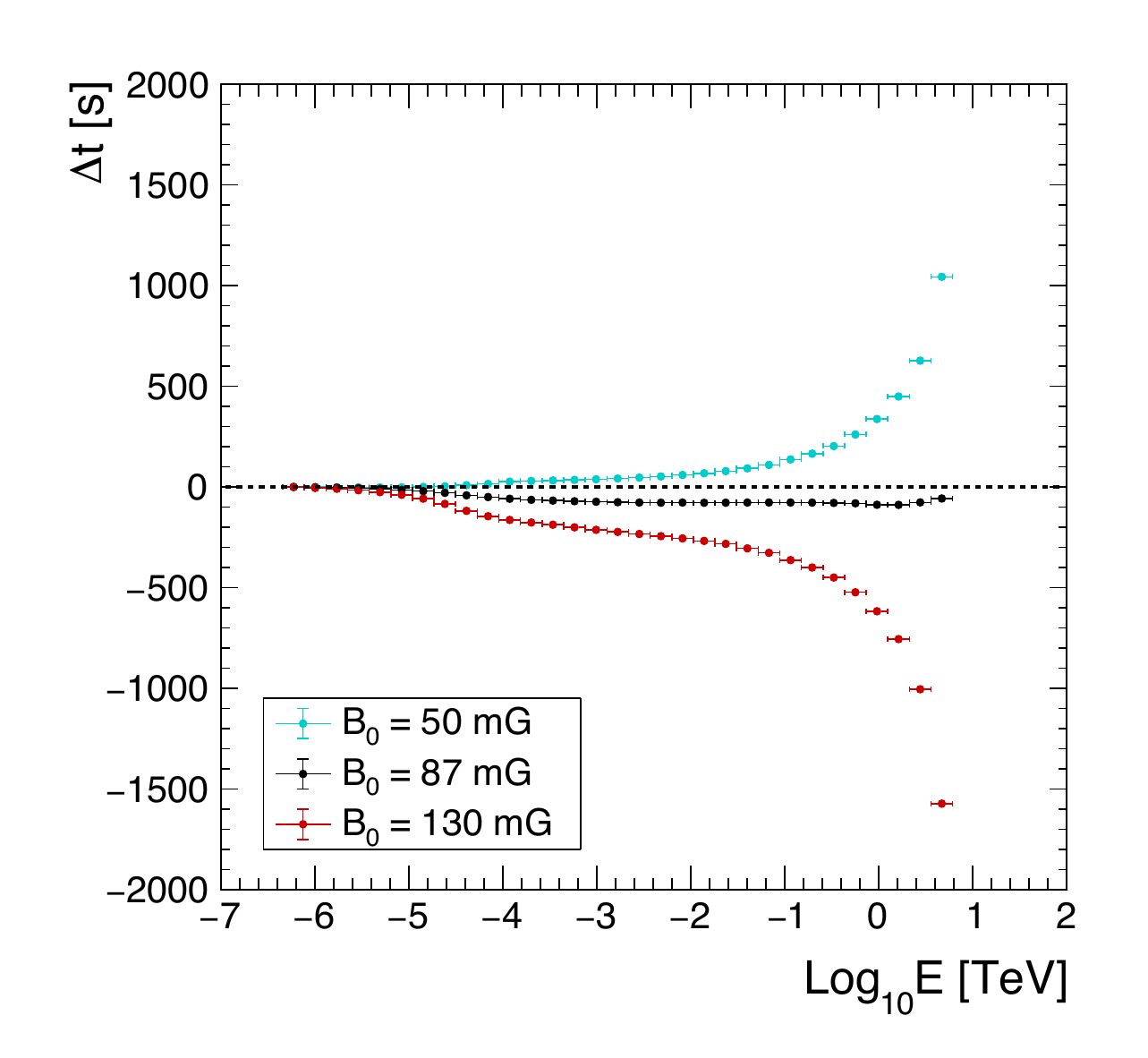}
  \caption{Intrinsic time delays at gamma-ray energies obtained for a pure SSC scenario, for the set of parameters given in Table~\ref{tab:parameters} (black curve, 87 mG) and for two other values of the magnetic field, 50 mG (blue) and 130 mG (red). The different trends (increasing in blue, flat in black, and decreasing in red) are here obtained only from the variation of the magnetic field amplitude $||\vec{B}||$, and start to appear at $E \sim 100$~MeV.}
              \label{fig:regimes}
    \end{figure}

\begin{table}
\caption{Set of parameters for the reference SSC scenario.\label{tab:parameters}}      
\label{table:1}      
\centering          
\begin{tabular}{l l l l } 
\hline
\hline          
SSC Parameters &  & Value & Unit \\ 
\hline
Redshift & z & 0.03 & - \\  
Doppler boost\tablefootmark{a} & $\delta_b$ & 40 & -\\
Magnetic field strength & $B_0$ & 87 & mG \\
Blob radius & $R_0$ & $5\times 10^{15}$ & cm  \\
Lepton density & $N_0$ & 300 & cm$^{-3}$ \\
Lorentz boost minimum & $\gamma_{min}$ & 2 & - \\
Lorentz boost cut-off & $\gamma_{cut}$ & $4\times 10^4$ & - \\
Power law index & n & 2.4 & - \\
\hline\hline         

Evolution Parameters &  & Value & Unit \\ 
\hline
Acceleration strength & $A_0$ & $4.5\times 10^{-5}$ & s$^{-1}$ \\
Acceleration evolution index & $m_a$ & 5.6 & - \\
Magnetic field evolution index& $m_b$ & 1 & - \\
Blob radius evolution index & $m_r$ & 0 & - \\
\hline
\end{tabular}
\tablefoot{
\tablefoottext{a}{Doppler boost $\delta_b$ depends on the bulk Lorentz factor of the blob, denoted $\Gamma_b$ in the main text, and on the angle between the jet axis and the line of sight.}
}
\end{table}

 \begin{figure}
  \centering
  \includegraphics[width=0.9\linewidth]{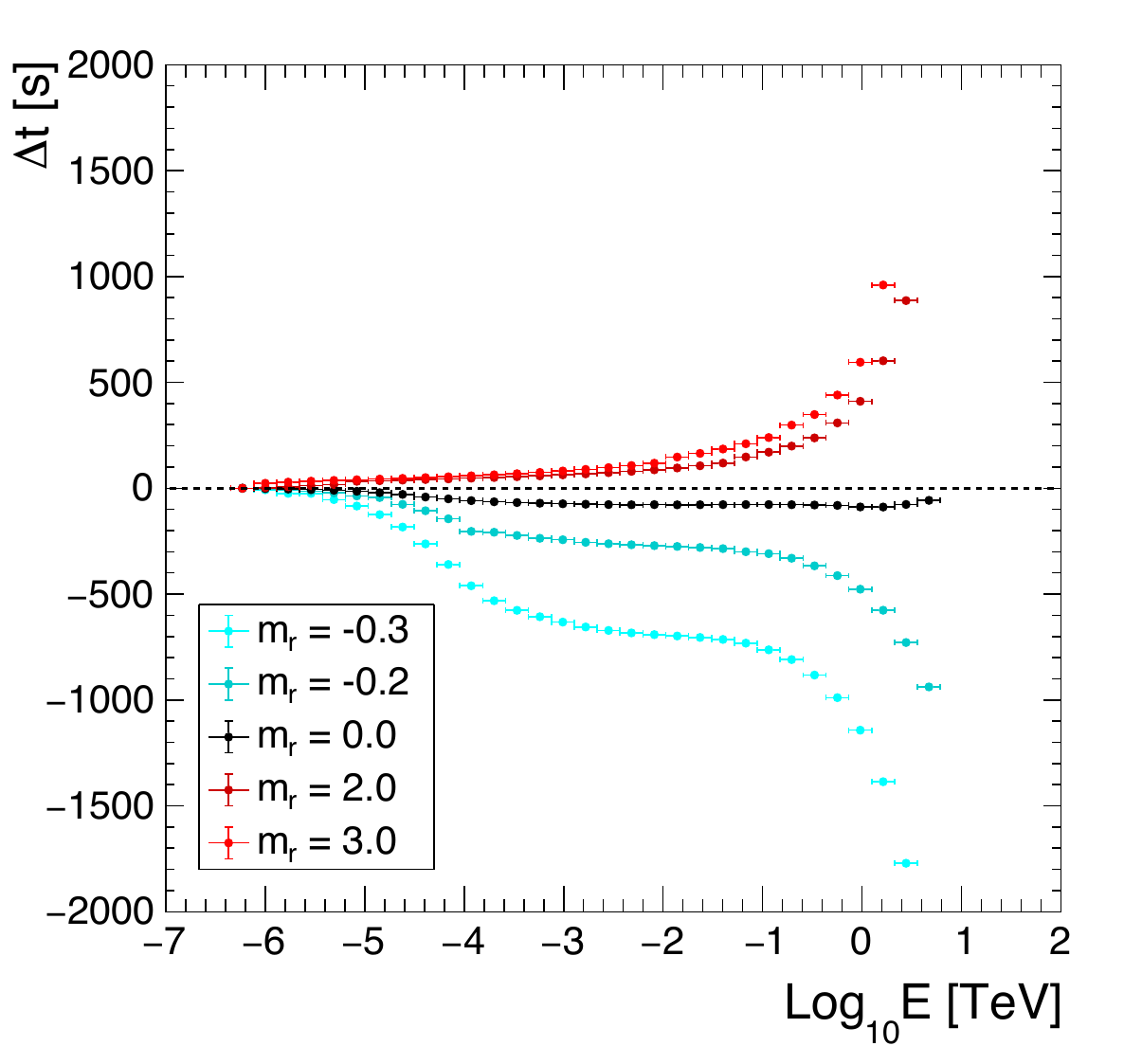}
  \includegraphics[width=0.9\linewidth]{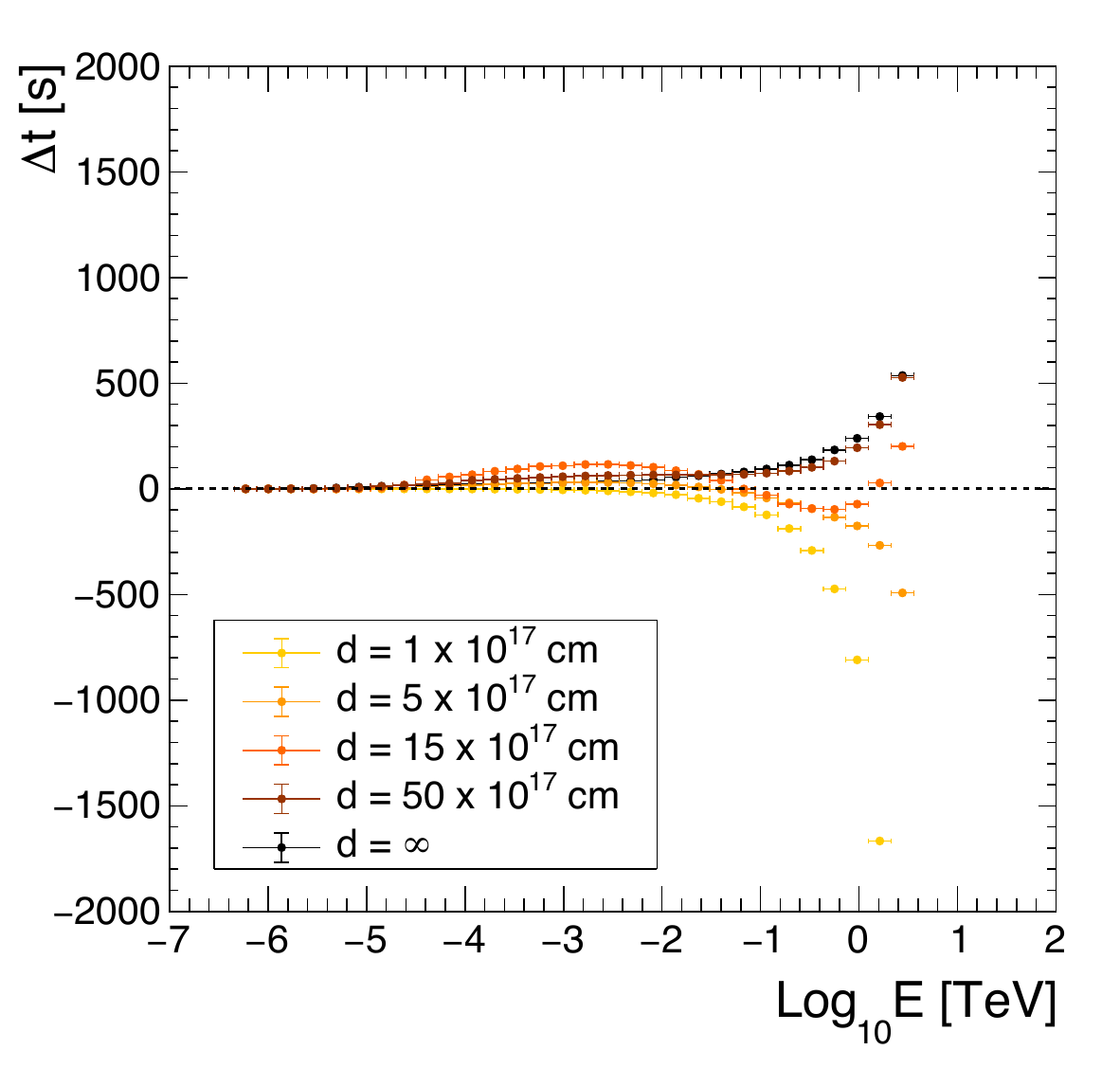}
  \caption{Intrinsic time delay as a function of the energy. Upper panel: Intrinsic time delay evolution for adiabatic expansion and compression effects, introduced with the variation of the blob's radius $R$. Lower panel: Illustration of delay evolution for decreasing external photon field effects, modelled as a blackbody emitted by an accretion disc, and further reprocessed by a BLR, varied by the increasing distance $d$ between the central black hole and the emitting blob. For the SSC part, the simulations are based on the set of parameters given in Table~\ref{tab:parameters}. For both plots, the reference time for lag calculation is taken at 1 MeV (see the main text for further explanations).}
    \label{fig:extended}
    \end{figure}

Each light curve can be characterised by a typical arrival time of photons, chosen to be the time at which the light curve peaks. The evolution of light curves over energy thus gives access to the evolution of the arrival times over energy. To estimate temporal delays as per energy band centred on energy $E$, we introduce a reference arrival time denoted $t_{\mathrm{ref}}$. The time delays are then given by
\begin{equation}
\label{eq:deltate}
\Delta t_{E} = t_{E} - t_{\mathrm{ref}}.
\end{equation}
We note the following:
\begin{itemize}
\item $\Delta t_{E}<0$ when the light curve centred on $E$ reaches its maximum before the reference light curve peaks;
\item $\Delta t_{E}>0$ when the light curve centred on $E$ reaches its maximum after the reference light curve peaks.
\end{itemize}
Since all light curves start at $t=0$, peaks necessarily occur at $t_{E}>0$ and the positiveness or negativeness of the delays is solely dependent on the chosen reference time. 

We are primarily interested in the time delay evolution with respect to energy. These intrinsic delays navigate between two types of regimes at high energy ($E>100$ MeV) in the pure SSC scenario: increasing and decreasing trends in the TeV range. These two regimes, respectively called long-lasting acceleration (or acceleration-driven), and fast acceleration (or cooling-driven) regimes in \cite{Perennes}, arise from an imbalance between acceleration and cooling processes. Leptons start to emit photons as soon as they are accelerated, such that the two processes take effect simultaneously, but with different relative influence. A transition zone between the two regimes, here referred to as flat trend, can arise from the combination and overall balance between the two processes. In this transition zone, intrinsic delays do not evolve and seem to cancel out, leading to a flat curve showing no variation with the photon energy. An example of intrinsic time delay evolution with energy can be found in Fig.~\ref{fig:regimes}, where $t_{\mathrm{ref}} = t_{E=1 \mathrm{MeV}}$. The three trends are represented, here obtained by varying the initial value of the magnetic field.

 In this work, and in order to investigate the impact of adiabatic and EIC processes on the intrinsic delays, we define a set of reference parameters permissible for a typical TeV-emitting blazar, with the intention of reproducing the special case of a flat trend (Table~\ref{tab:parameters}). This choice reflects the fact that no significant time delay at TeV energies in AGN has been detected yet, with the notable exception of the 4$\pm$1 min delay detected by MAGIC in 2005 for the archetypal BL Lac object \object{Mkn 501} at redshift $z = 0.03$ \citep{4min}. The flat regime therefore appears to be a better fit to observations. This state is fairly unstable and slight variations of parameters can send delays along an increasing or decreasing trend, and can thus be used to better highlight the impact of individual model parameters on intrinsic delays.

 The impact of adiabatic processes is illustrated in Fig.~\ref{fig:extended} (upper panel), which shows the variation of intrinsic time delays with respect to the blob's radius evolution index $m_r$. The reference value $m_r = 0$ (the radius stays constant) is represented by the black curve. The red points correspond to expansion of the blob ($m_r > 0$), while the blue points correspond to a compression ($m_r < 0$). When the blob expands with time, as expected in the absence of an efficient confinement mechanism, its plasma is diluted and leptons diffuse their energy. This leads to a non-radiative additional cooling effect placing delays in the long lasting acceleration regime with an increasing trend. On the contrary, if the blob is compressed during the flare and its volume decreases with time, its plasma becomes more condensed and is heated up. This rise in energy translates into a gain in kinetic energy, and leptons are naturally accelerated, placing delays in the fast acceleration regime with a decreasing trend. Thus, it appears adiabatic processes do not modify the overall description of intrinsic time delays and only influence the regime in which the delays evolve with energy for a specific given flare.

The evolution of intrinsic time delays with respect to an external photon field contribution is shown in Fig.~\ref{fig:extended} (lower panel). All the parameters governing the external field are kept constant with the exception of the blob-accretion disc distance $d$.
In this configuration with non-zero EIC, numerical computations can diverge when the adiabatic expansion index $m_r$ is smaller than or equal to 1 \citep{Levy}. For that reason, we adopt the value $m_r = 1.001$. The other parameters are set to the values given in Table~\ref{tab:parameters}. At very large distances $d$, the case $d = \infty$ is equivalent to a vanishing EIC contribution and the lag evolution follows an increasing trend due to the adiabatic expansion effect. In our description, reducing the distance $d$ corresponds to increasing the energy density of the external photon field and the importance of the EIC radiation and cooling (see Eq.~\ref{eq:EIC}). The presence of additional EIC simply increases the radiative cooling without modifying the acceleration term.
This induces a global evolution of the intrinsic delays and pushes them towards the decreasing trend at VHE. 
The situation is a bit more complex at intermediate distances ($d = 15 \times 10^{17}$~cm) and intermediate gamma-ray energies where delays first increase for $E<1$~GeV, decrease between $E=1$~GeV and $E=1$~TeV, and increase again for $E>1$~TeV. 
This effect results from the combination of the three different cooling mechanisms (SSC, EIC, and adiabatic expansion), which are more or less important depending on the energy, relative to the acceleration process. However, this behaviour does not call into question the presence of the two main regimes already identified, with increasing or decreasing trends, which characterise the intrinsic time delays for general leptonic scenarios at TeV energies and above, where LIV effects could be expected.

We can conclude that the presence of adiabatic processes and of a significant EIC emission do not modify the overall picture of intrinsic leptonic time delays in the gamma-ray range, so FSRQs should follow intrinsic patterns similar to those of BL Lacs. However, it can significantly modify the regime of the intrinsic delays. Significant EIC ultimately pushes intrinsic delays towards the decreasing trend (cooling-driven regime) at VHE, and adiabatic compression does the same, while adiabatic expansion pushes intrinsic delays towards the increasing trend (long-lasting acceleration regime). The model described here yields intrinsic delays quite effortlessly from the variation in source physical quantities and properties. A variation of about 50\% in the value of the main model parameters can indeed be enough to induce intrinsic time delays larger than 200~s/TeV (as shown in Fig.~\ref{fig:regimes} for the magnetic field), which could be potentially detectable, depending on the experimental errors. At a first glance, this might appear to be in contradiction with the lack of detected time delays at TeV energies blazar data collected so far. However, such a non-detection is usually explained by a lack of precision of the current measurements, so that the uncertainty on time lag measurements is still too large, leading to values compatible with zero. This question should be directly answered by the next generation of Cherenkov instruments, such as the Cherenkov Telescope Array \citep{Rosales2024}. If no time delay were ever detected at VHE gamma rays, this would impose strong constraints on standard leptonic flare models and could put them in difficulty and even discard some of them.

\section{Exploring multi-wavelength leptonic flares with an additional Lorentz invariance violation delay}\label{sec:addingliv}

\subsection{Injection of Lorentz invariance violation delays}

In order to properly compare LIV and intrinsic time delay characteristics, the LIV delay computation was implemented in the simulation code. Each light curve is shifted by a value following
\begin{equation}
\label{eq:int+LIV}
t \longrightarrow t + \tau\, E_\mathrm{LC},
\end{equation}
where $E_\mathrm{LC}$ is the mean light curve energy and $\tau$ is the linear LIV term (Eq.~\ref{eq:tau}). It should be noted that this transformation does not change the shape of individual light curves, but it modifies total time delays as well as the temporal evolution of the SEDs, which therefore have to be rebuilt from the light curves in order to include the LIV effect. The parameter
$\tau$ is kept as a free parameter expressed in s\,TeV$^{-1}$, and it can take positive (subluminal LIV effect) or negative (superluminal LIV effect) values. A typical uncertainty on measured parameter $\tau$ equivalent to a $1 \sigma$ confidence level for \object{Mkn 501} is $\sim200$ s\,TeV$^{-1}$ \citep[][Table A1]{HMV}, and so it was chosen as a typical value for the injected lag.

\begin{figure}
  \centering
  \includegraphics[width=0.9\linewidth]{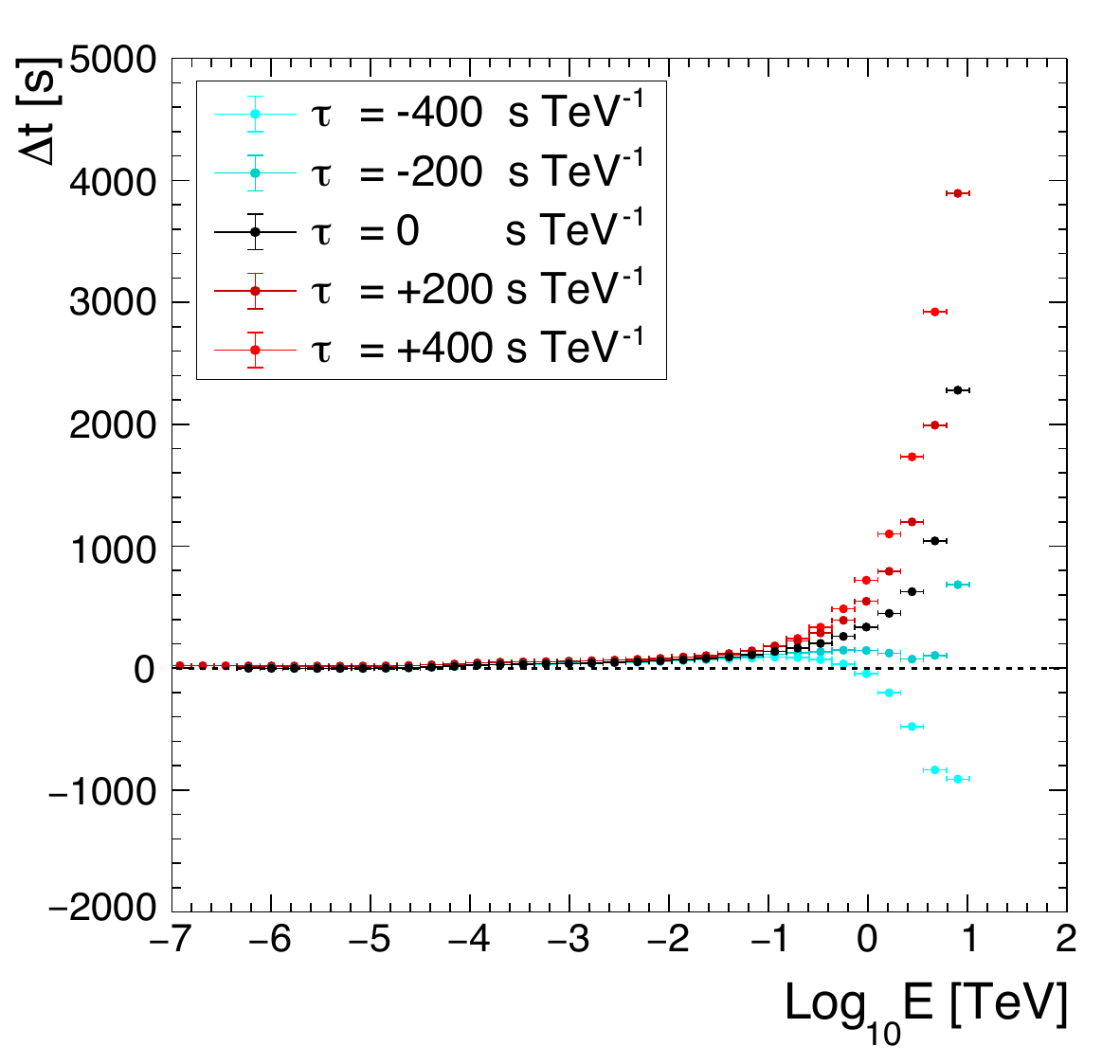}
  \caption{Simultaneous description of intrinsic and LIV effects in time delays at high energies, illustrated by varying the importance of LIV around the case of the intrinsic increasing trend from Fig.~\ref{fig:regimes} with $B_0 = 50$~mG (in black). Positive (superluminal) and negative (subluminal) values of $\tau$ are represented in red and blue, respectively.}
              \label{fig:LIV}
    \end{figure}

With such considerations, total time delays observed for a flare with an intrinsic increasing trend and for various LIV contributions can be found in Fig.~\ref{fig:LIV}. Intrinsic effects are shown in black, while intrinsic combined with LIV effects are shown in red for $\tau > 0$ and blue for $\tau < 0$. As LIV contribution depends on energy, having $\tau > 0$ tends to send delays towards the increasing trend, while $\tau < 0$ sends them towards the decreasing trend. Therefore, LIV either amplifies time delays when both effects impose the same evolution with energy, or tends to suppress time delays when the two effects impose opposite evolution. For $\tau$ that is large enough, LIV effects can even change the observed trend at high energies. Then, it appears that separating LIV-induced and intrinsic delays can be challenging without a specific strategy.

\subsection{Multi-wavelength X-ray and $\gamma$-ray flares from blazars}

The study of intrinsic time delays performed in Sect.~\ref{sec:intdelays} focuses on the gamma-ray domain. This limitation was motivated by the fact LIV effects can only become significant at very high energies. However, intrinsic delays also arise from the low-energy emission as discussed for instance by \cite{Lewis}.
Within leptonic scenarios, the inverse-Compton high-energy bump is directly dependent on the synchrotron low-energy bump. A strong correlation between the two bumps of the SEDs is expected since it is the same population of high-energy leptons that produces the synchrotron emission and the high-energy emission due to the inverse-Compton process. The correlation is even greater for pure SSC models where the ambient scattered photons are the synchrotron photons emitted themselves by the same high-energy leptons. In any case, the two bumps are closely related and evolve together in a similar fashion for leptonic scenarios. This similarity between energy bands should then also appear in light curves and more importantly in intrinsic time delays. Since LIV effects should only appear in the high-energy domain, studying the relationship between sets of intrinsic delays arising in low- and high-energy ranges could provide strategies based on multi-wavelength observations to discriminate between intrinsic and LIV-induced time delays. The occurrence of LIV effects should result in breaking or weakening the similarity between observed low-energy and high-energy time delays.

\begin{figure}
  \centering
  \includegraphics[width=0.9\linewidth]{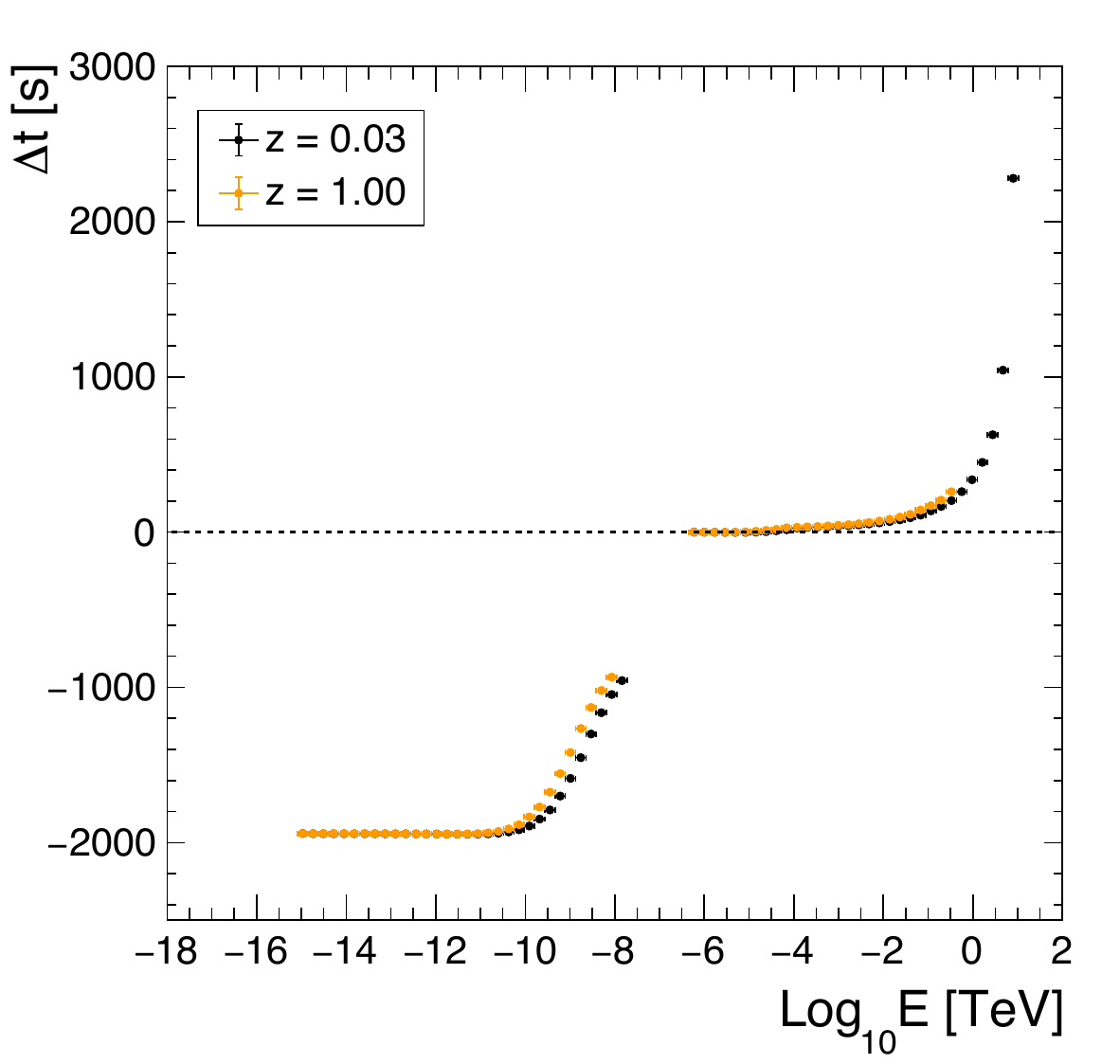}

  \caption{Intrinsic time delays extended to low-energy domains for the case of the intrinsic increasing trend from Fig.~\ref{fig:regimes}, obtained for low redshift (z = 0.03, black) and high redshift (z = 1, orange). Here the energy reference value is taken at $E_{\mathrm{ref}}=1$ MeV. At high redshift, fluxes at the highest energies are reduced and the corresponding data points are suppressed due to strong EBL absorption. The regime found at low redshift is retained nonetheless. No LIV delay is injected here.}
              \label{fig:Z}
    \end{figure}

To test this hypothesis, we extended the energy range to derive intrinsic delays at X-ray, optical, and infrared energies.
To better separate low- and high-energy domains, special care was given to the choice of the reference arrival time $t_{\mathrm{ref}}= t_{E = E_{\mathrm{ref}}}$ corresponding to a specific energy $E_{\mathrm{ref}}$ (Eq.~\ref{eq:deltate}). The energy of the dip in SEDs, the low-flux region between the two bumps, was chosen for this quantity. Since the location of the dip varies with the time evolution of the generated SEDs, the SED with the highest $\nu F_{\nu}$ flux value at that location was considered as a reference. The energy $E_{\mathrm{ref}}$ is then taken as the energy of the dip for that specific SED. The value of $E_{\mathrm{ref}}$ typically varies between 0.01 and 1 MeV for BL Lacs. To highlight the relationship between the two energy domains, the parameter space and the different regimes of delays were explored.
One illustrative example is shown in Fig.~\ref{fig:Z} for a typical flare of a BL Lac object at low and high redshifts. 
A clear similarity between the two energy domains (synchrotron and inverse-Compton) appears in the absence of any LIV effects.
This behaviour was found to be systematic: both sets of low- and high-energy delays always follow the same trend whatever the set of parameters used to generate them.
This systematic similarity indicates that the Klein-Nishina and EBL effects, which tend to modify the observed fluxes at the highest energies, do not significantly affect the intrinsic delays, and by extension their regime of evolution with energy. In the case of \mbox{FSRQs}, the symmetry between synchrotron and inverse-Compton domains remains, but is weakened compared to the pure SSC case, due to the introduction of an external photon field. The external inverse-Compton emission then gains an additional component independent from the synchrotron component, which somewhat reduces the inter-dependency between the delays in the X-ray and gamma-ray domains.

At high redshift $(z=1)$, observed fluxes become much smaller for the same set of source parameters, and the EBL contribution is much stronger, and hence suppresses the very high-energy tail of the SEDs. 
In order to be compatible with CTA sensitivity \citep{CTAperf}, a selection was made where a threshold is applied to the observable flux. Only datasets with flux $F>10^{-20}\ \mathrm{erg\,cm^{-2}\,s^{-1}}$ are used in our analysis. As a consequence, modifying the redshift changes the selection on the datasets, as
can be seen in Fig.~\ref{fig:Z} where the time delay evolution (increasing regime) with redshift $z=1$ (orange points) is compared to that with the reference redshift $z=0.03$ (black points). While a high redshift simply eliminates information at VHE, and thus removes the data points, the regime and shape of the distributions are preserved. The values of the delays change only slightly in the X-ray and the gamma-ray domains. The same qualitative results hold independently of the regime of the intrinsic time delays. This illustrates the weak dependency between intrinsic time delays and redshift.

\subsection{Disentangling intrinsic delays from LIV effects}

\subsubsection{Euclidean distance technique}

The first method proposed to disentangle intrinsic and LIV delays is based on the computation of the Euclidean distance between the delay versus energy curves at low and high energies. It quantifies the degree of similarity between the shape of two distinct datasets, here the time delays in the X-ray and the VHE gamma-ray domains.
The Euclidean distance between the two datasets $A[i]_{(1 \leqslant i \leqslant n)}$ and $B[i]_{(1 \leqslant i \leqslant n)}$ containing $n$ points is simply written as 
\begin{equation}
\label{eq:eucl_scalar}
d_{E} = \frac{\sqrt{\sum_i \left(A[i]-B[i]\right)^2}}{\sqrt{\sum_i \left(A[i]+B[i]\right)^2}},
\end{equation}
where the denominator is a normalisation term, which is introduced for better graphical representation and readability. The scalar $d_{E}$ possesses two specific identities:
\begin{itemize}
\item $d_{E}=0$, which is the minimum value, indicating a perfect match between the two datasets, that is $A[i] = B[i], \forall i \in [1,n]$;
\item $d_{E}=1$, which indicates that $\forall i \in [1,n]$, we have $A[i]=0$ or $B[i]=0$.
\end{itemize}
For this multi-wavelength study, two datasets are generated. The first contains synchrotron delays (with inverse-Compton delays set to zero) and the other contains inverse-Compton delays (with synchrotron delays set to zero). In other words, $A[i]$ (respectively $B[i]$) corresponds to the values of $\Delta t$ below (respectively above) the dip in the SED.
To properly apply the method, we make sure there is the same number of non-zero points in each dataset, a necessary condition for the method to work as intended and steer relevant information. Then, the two sets of delays are extended between $\log_{10}E = -16$ and $\log_{10}E = 2$ ($E$ being expressed in TeV), filling the areas where there is no information with zeros. To maximise the sensitivity, we re-scale all the data points with the first non-zero delay value $\Delta t_{\mathrm{first}}$: $\Delta t \longrightarrow \Delta t - \Delta t_{\mathrm{first}}$.
Finally, we also need to shift one of the datasets over energy towards the other one to provoke their overlapping. This allows us to find the optimal displacement that will minimise the Euclidean distance $d_{E}$. To estimate this optimal displacement, we introduce a modified version of Equation \ref{eq:eucl_scalar}, where $d_{E}$ now becomes a function of the displacement $k$,
\begin{equation}
\label{eq:eucl}
d_{E}(k) = \frac{\sqrt{\sum_i \left(A[i-k]-B[i]\right)^2}}{\sqrt{\sum_i \left(A[i-k]+B[i]\right)^2}},
\end{equation}
with $A$ the displaced dataset. For the present study, we computed the distance in logarithmic scale by displacing the synchrotron dataset ($\equiv A$) over energy by a quantity $\epsilon = 10^k$ towards the inverse Compton set ($\equiv B$) such that $E_{\mathrm{new}} = E\times 10^k$, with $k$ varying from 0 to 16. Although our datasets are discrete, we treated the distributions of time delays as functions and retrieved values with linear interpolations between consecutive points such that $k$ can take any real value.

\subsubsection*{For $\tau = 0$: intrinsic effects only}

\begin{figure}
  \centering
  \includegraphics[width=0.9\linewidth]{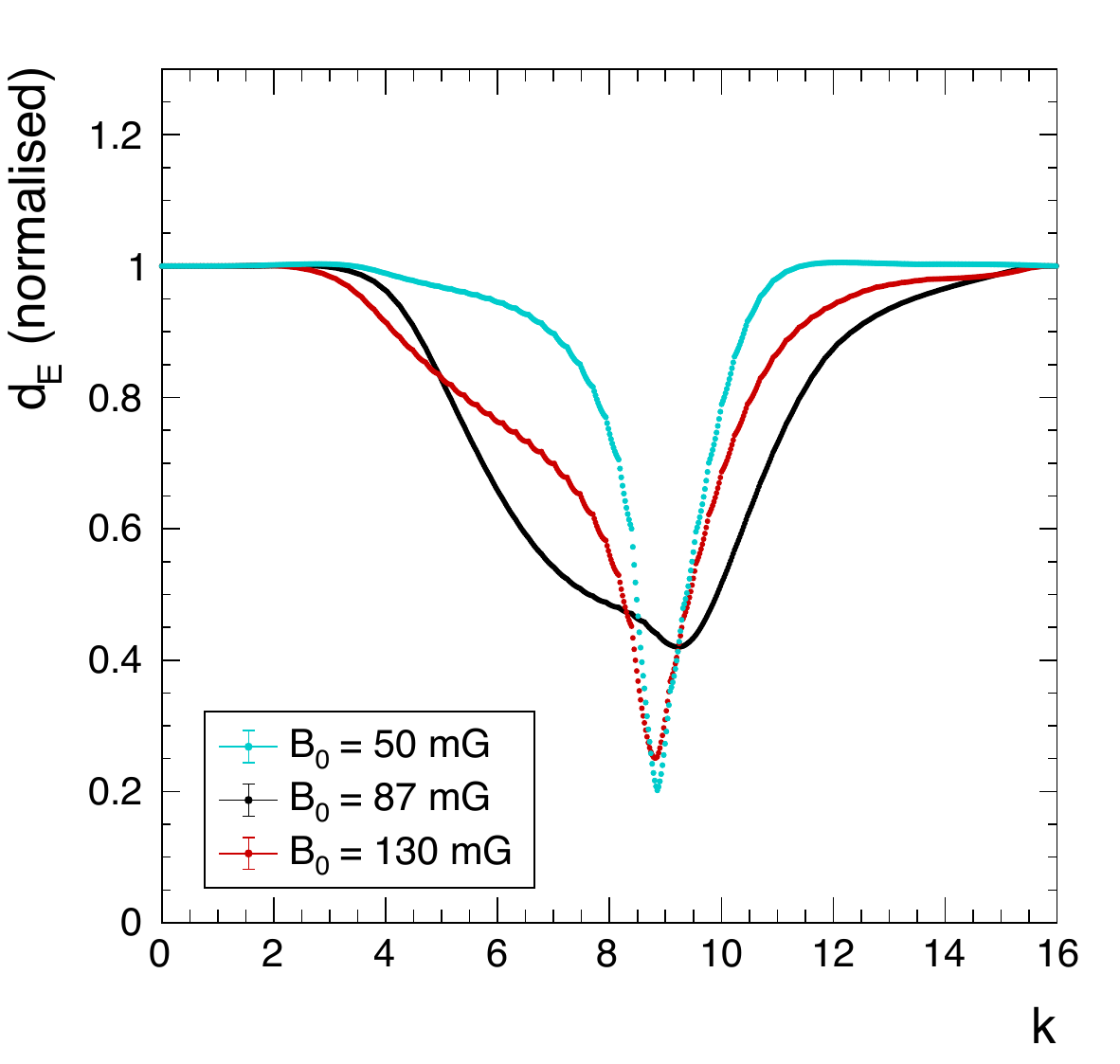}
  \caption{Euclidean distance functions computed between intrinsic time delays from the synchrotron and inverse-Compton domains for the three cases presented in Fig.~\ref{fig:regimes}. Each case yields a clear minimum defining the optimal displacement $k_{\mathrm{min}}$ for which we obtain the best match between the two sets of delays. Blue: increasing trend. Black: flat trend. Red: decreasing trend.}
              \label{fig:ED}
    \end{figure}

The Euclidean distance between the synchrotron and inverse Compton datasets is shown in Fig.~\ref{fig:ED} for three flares illustrating the different intrinsic delay regimes. 
In all cases (increasing, flat, and decreasing trends), the figure shows that the Euclidean distance $d_{E}$ reaches a clear minimum $d_{E,\mathrm{min}} = d_E(k_\mathrm{min})$ for intermediate values of $k$, at $k_\mathrm{min}$ typically of the order of 9, corresponding to the best match between the shape of the two sets of delays. For small $k$ (here $k < 3$), $d_{E}=1$ because the non-zero values of the two sets of delays do not yet overlap. As $k$ increases, datasets of non-zero values start to overlap, and the Euclidean distance gradually shifts away from 1 and towards its minimum value. Once $k$ increases above $k_\mathrm{min}$, the similarity between the two datasets and their overlapping area start to decrease again and the Euclidean distance slowly tends towards $d_{E} = 1$.
We also note that the specific case of flat trend (in black), which presents little variation of the delays with the frequency, tends to reduce the sensitivity on the Euclidean distance method, reaching a smooth minimum of $d_{E,\mathrm{min}} \sim 0.42$ at $k \sim 9.2 \equiv k_{\mathrm{min}}$. Euclidean distance functions for the decreasing and increasing trends reach significantly lower values ($d_{E,\mathrm{min}} \sim 0.25$ and $d_{E,\mathrm{min}} \sim 0.20$ respectively) than for the flat trend case. This is due to the amplified variation of time delays with the energy, leading to a minimum in the Euclidean distance function, which is better defined.

All flares with intrinsic time delays significantly increasing or decreasing at high energies show a small and well-defined minimum Euclidean distance $d_{E,\mathrm{min}}$, however, without reaching a perfect match ($d_{E,\mathrm{min}} = 0$) between their synchrotron and inverse-Compton delays. Simulations of several flares covering a large space of parameters extended around the reference set of Table~\ref{tab:parameters} show that the optimal displacement $k$ is always such that $8 < k_\mathrm{min} < 10$ and that the minimum Euclidean distance is generally $d_{E,\mathrm{min}} < 0.5$ and always such that $d_{E,\mathrm{min}} < 0.6$. The maximum value of 0.6 is reached for some flares with time delays following a flat trend.

\subsubsection*{For $\tau \neq 0$: intrinsic and LIV effects}

\begin{figure*}
  \centering
 \includegraphics[width=0.95\linewidth]{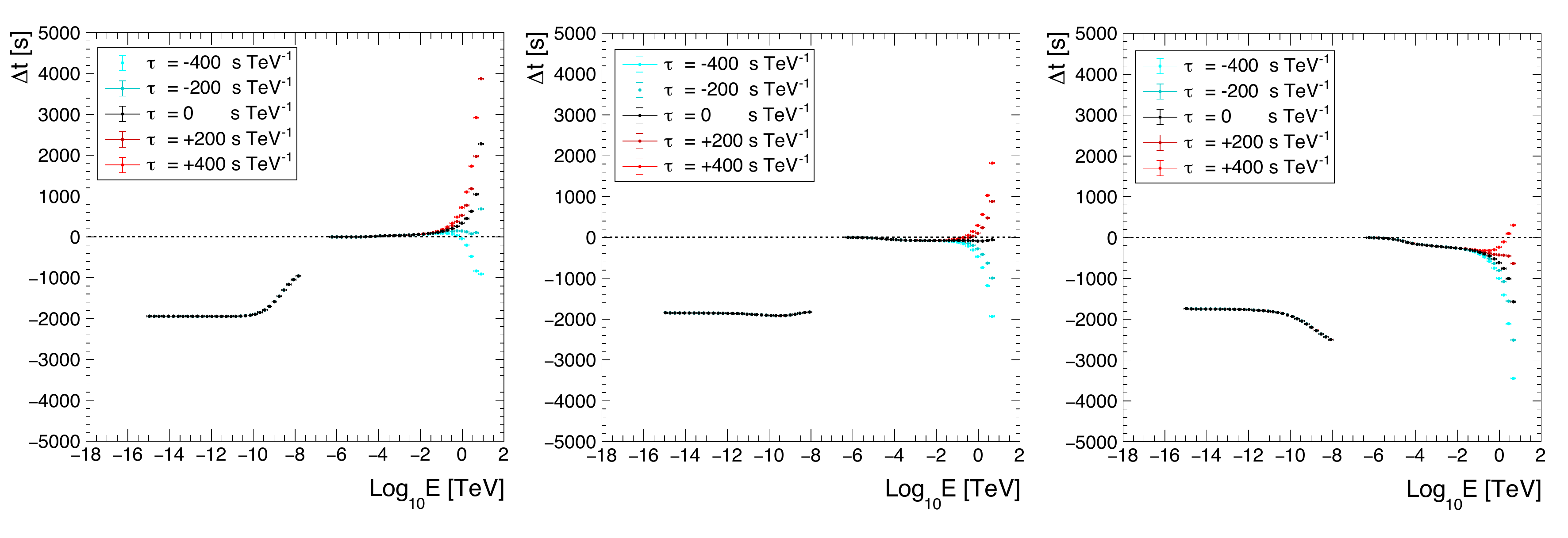}
 \includegraphics[width=0.95\linewidth]{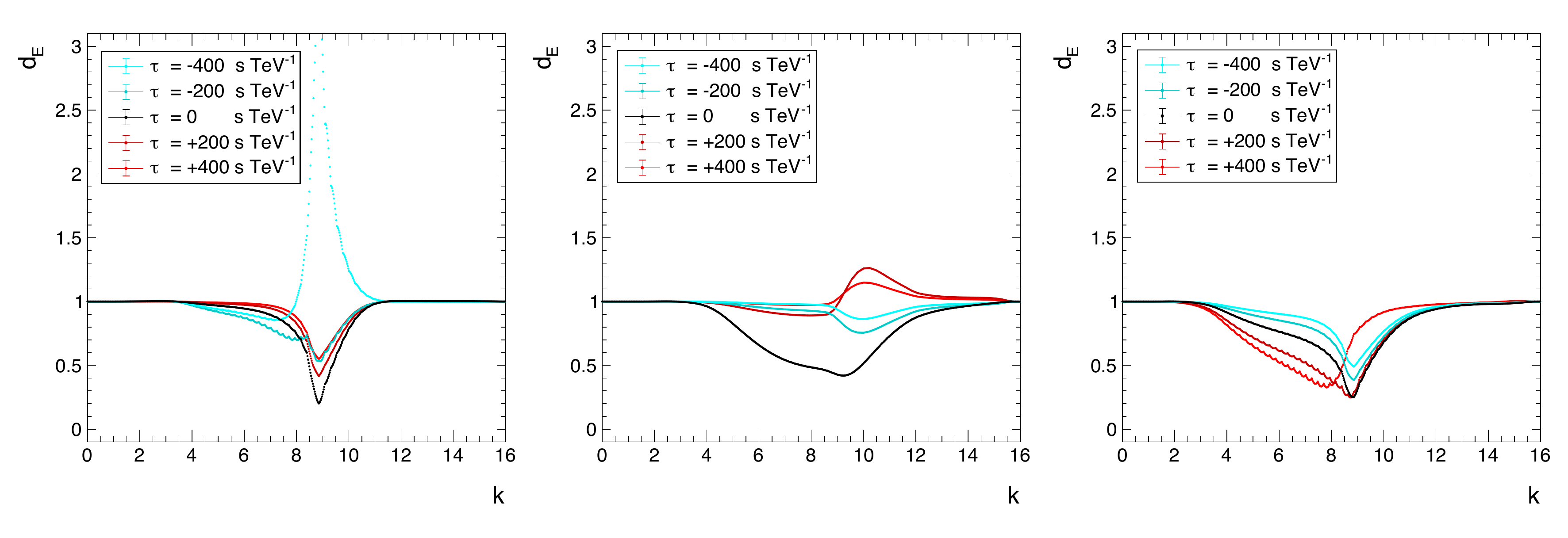}
  \caption{Time delays (upper panel) and Euclidean distance function (lower panel) computed for the three flares of Fig.~\ref{fig:regimes} with various LIV contributions. The minimum Euclidean distance $d_{E,\mathrm{min}}$ is always reached by the curve with no LIV effect (black). Left: Increasing trend at $\tau = 0$; centre: Flat trend; right: Decreasing trend.}
              \label{fig:LIVED}
    \end{figure*}

Since LIV mostly affects the inverse-Compton domain at very high energies, while leaving the synchrotron domain unaltered, its contributions tend to decorrelate synchrotron and inverse-Compton sets of time delays. As a consequence, the presence of LIV delays can significantly modify the initial pattern of the Euclidean distance function. As illustrated by several cases shown in Fig.~\ref{fig:LIVED}, the distance function minimum $d_{E,\mathrm{min}}$ 
can drastically increase when $\tau$ becomes different from 0, and the optimal displacement $k_{\mathrm{min}}$ can be modified as a consequence of highly decorrelated sets of data. 
There are, however, some specific cases where 
the introduction of LIV leaves the minimum Euclidean distance almost unchanged, as illustrated by a few cases in Fig.~\ref{fig:LIVED} (e.g. bottom row, right panel, for $\tau = +200$~s\,TeV$^{-1}$). Nevertheless, as the introduction of LIV tends to decorrelate the time delays, it always yields either larger or equal values of $d_{E,\mathrm{min}}$.

Assuming that the next generation of instruments will be able to observe bright flares with a good temporal and large spectral coverage and to detect non-zero time delays, then the Euclidean distance function will become an observable. Fitting it with a hybrid (leptonic + LIV) model should make it possible, in favourable cases, to discriminate between the two origins of the delays. However this will require very high-quality sets of data. As a first step, simpler criteria could be applied, for instance taking advantage of the constraints we found above on $d_{E,\mathrm{min}}$ and $k_\mathrm{min}$ (for $\tau = 0)$. The detection of some $d_{E,\mathrm{min}} > 0.6$, $k_\mathrm{min} < 8$, or $k_\mathrm{min} > 10$ coming from an intrinsic SSC flare would emphasise the presence of additional non-intrinsic delays possibly due to LIV. 
 
\subsubsection{Hardness-intensity diagrams and hysteresis patterns}

The degree of similarity between synchrotron and inverse-Compton domains can also be quantified directly from SEDs. Hardness-intensity diagrams (HIDs), showing the time evolution of the SED index (hardness) as a function of the SED flux (intensity), can give rise to specific hysteresis patterns. Such patterns have been found in observed data \citep[e.g. in][]{Hyst_obs2,Hyst_obs3,Hyst_obs1}, but also simulated with AGN models \citep[e.g. in][]{Hyst_mod}. They should ultimately help to constrain the emission scenarios. The study of hysteresis patterns appears to be complementary to the Euclidean distance technique and can help consolidate this approach for discrimination between intrinsic and LIV-induced effects in time delays.

\subsubsection*{For $\tau = 0$: intrinsic effects only}

To produce HIDs, the hardness is computed over two small energy windows that need to be the same for all SEDs, one in the synchrotron bump and the other in the inverse-Compton bump. For the purpose of the present study, the inverse-Compton window should be at energies as high as possible. The intensity plotted in the diagram is taken as the mean flux within the energy window, which should be relatively small for the intensity to vary as little as possible. Finally, the window should not overlap with the bumps' maxima to prevent a change in sign of the spectral index. We therefore define the inverse-Compton (resp. synchrotron) window as follows: the lower boundary is set to the position of the inverse-Compton (resp. synchrotron) peak at the highest energy reached during the flare, and the width is set to one decade in energy.

  \begin{figure*}
  \centering
  \includegraphics[width=0.95\linewidth]{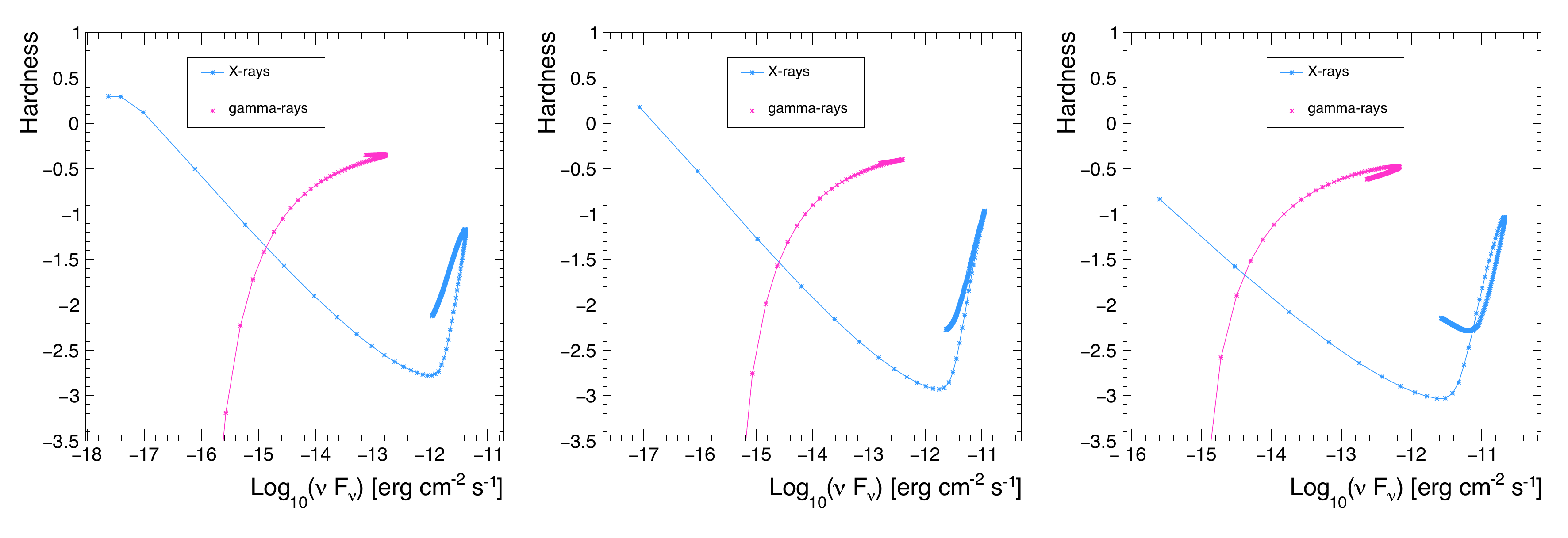}
  \caption{Hardness-intensity diagrams (HIDs) highlighting hysteresis patterns for X-ray (blue) and gamma-ray (magenta) domains for intrinsic effects only, namely for the three cases with $\tau = 0$ shown in Fig.~\ref{fig:LIVED}. Left: Increasing trend, anticlockwise hysteresis loop at high fluxes. Centre: Flat trend, almost no hysteresis. Right: Decreasing trend, clockwise hysteresis loop at high fluxes.}
              \label{fig:Hyst}
    \end{figure*}

The hysteresis pattern for the three flares of Fig.~\ref{fig:LIVED} illustrating the different trends of intrinsic delays can be found in Fig.~\ref{fig:Hyst}. 
The X-ray domain hysteresis loop is much wider than the gamma-ray one, as expected since the flux amplitude variation in the SED is more important in the synchrotron domain than in the inverse-Compton domain. However, the
X-ray and gamma-ray hysteresis always follow the same loop orientation at high fluxes.
The loop orientation depends on the nature of the flare, and changes at high fluxes according to its intrinsic delay regime. Flares with decreasing (resp. increasing) intrinsic trend show a clockwise (resp. anticlockwise) hysteresis loop at high X-ray and gamma-ray fluxes. Flares with a flat intrinsic trend show almost no hysteresis, the rising and decaying parts of the flare following the same paths on the HID at high fluxes, as can be expected when there is no intrinsic time delay.
Observing a hysteresis pattern at high flux in a dataset means there should also be non-zero intrinsic time delays in the SSC scenario; therefore, information on the X-ray domain hysteresis can predict the existence or the absence of intrinsic time delays in the VHE gamma-ray domain in the frame of standard physics.

\subsubsection*{For $\tau \neq 0$: intrinsic and LIV effects}

\begin{figure}
  \centering
  \includegraphics[width=0.9\linewidth]{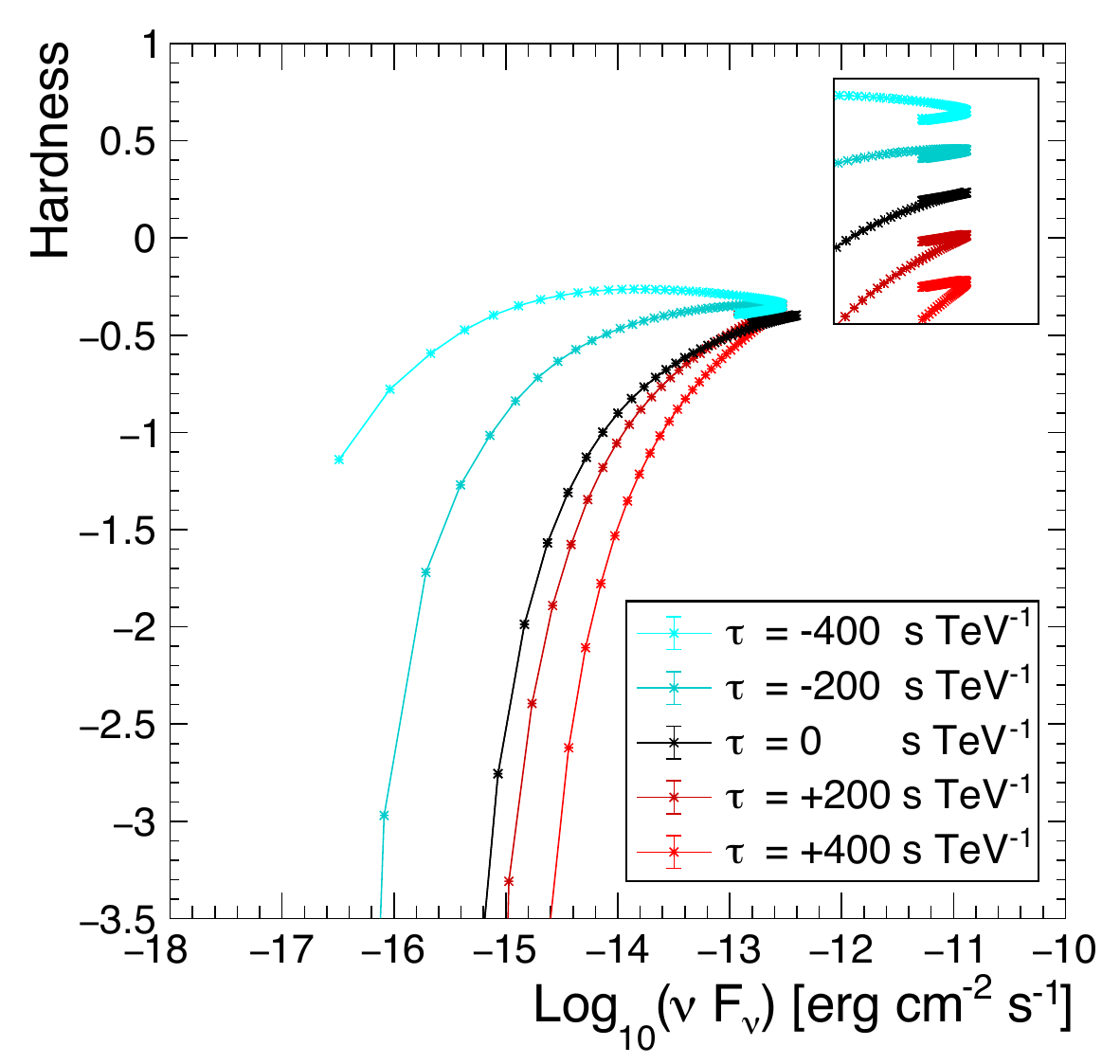}
  \caption{HIDs in the gamma-ray energy domain for the flat trend case of Fig.~\ref{fig:Hyst} (centre), with various additional LIV contributions (in colour). In the inset the curves are shifted, emphasising the change in curvature orientation at high flux values.}
              \label{fig:LIVHyst1}
    \end{figure}

\begin{figure}
  \centering
  \includegraphics[width=0.9\linewidth]{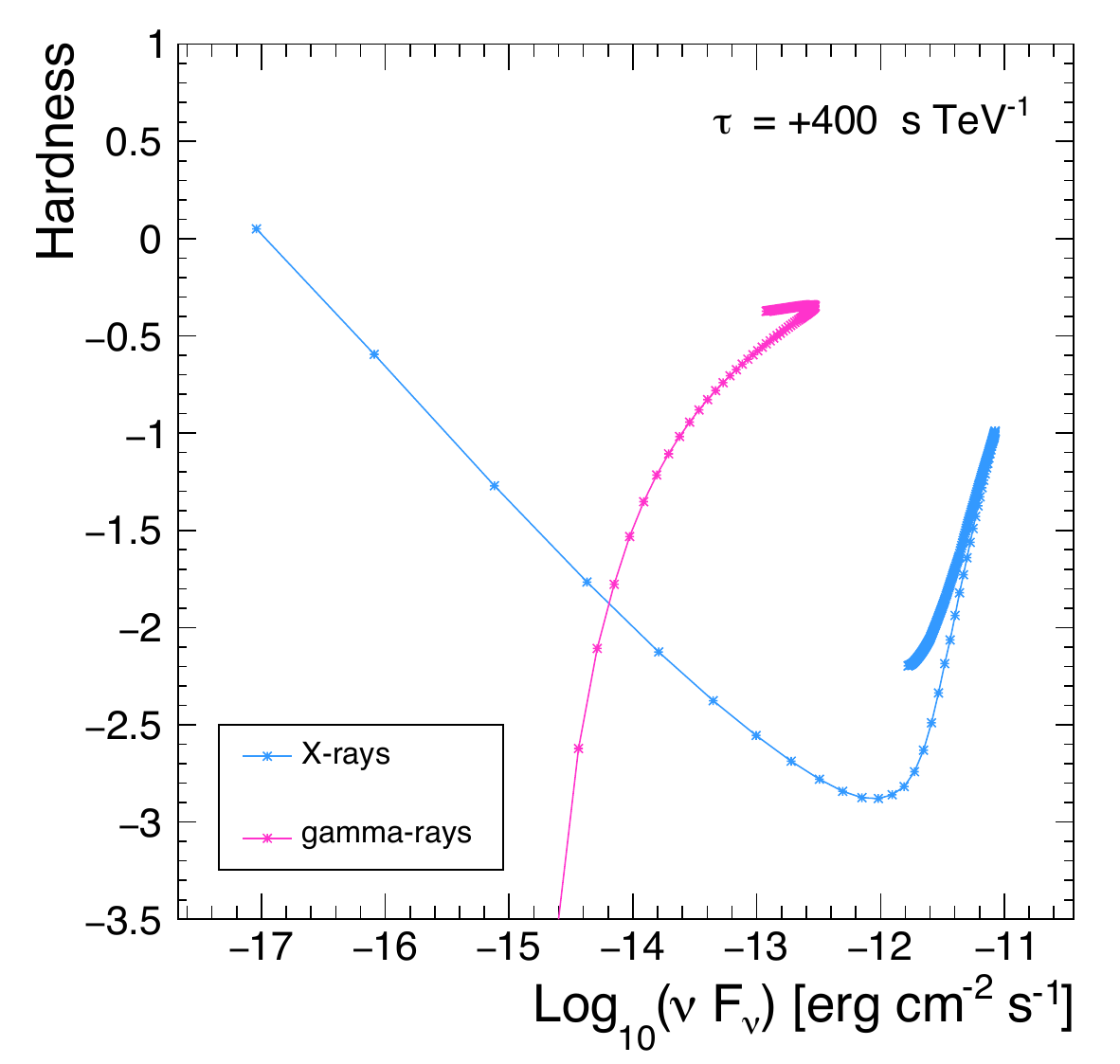}
  \includegraphics[width=0.9\linewidth]{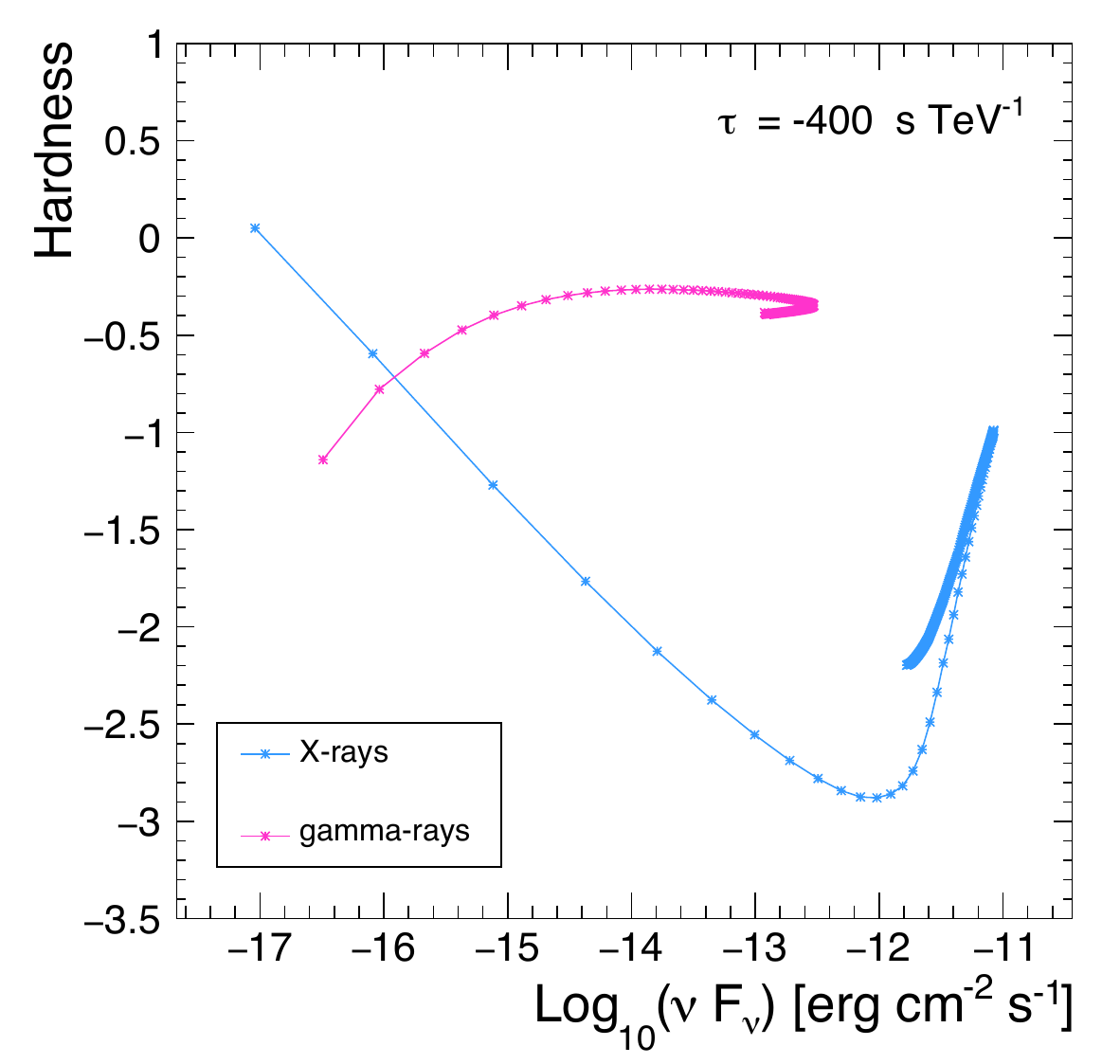}
  \caption{HID highlighting hysteresis patterns for X-ray (blue) and gamma-ray (magenta) domains for the flat trend case shown in Fig.~\ref{fig:Hyst} (centre) with additional LIV effects, namely $\tau = + 400$ s\,TeV$^{-1}$ (top pannel) and $\tau = - 400$ s\,TeV$^{-1}$ (bottom pannel), showing that LIV effects can significantly modify the hysteresis pattern and even cause a change in loop orientation in the gamma-ray domain. While the X-ray hysteresis pattern is left unchanged relative to the case at $\tau = 0$ in Fig.~\ref{fig:Hyst}, a rotation pattern appears in the gamma-ray pattern (either clockwise or anticlockwise depending on the sign of $\tau$).}
              \label{fig:LIVHyst2}
    \end{figure}

Contrary to time delays, the LIV contribution to the spectra is barely noticeable and cannot be resolved directly from observed SEDs. However, as hysteresis patterns are very sensitive to any variation in SEDs, modification due to LIV can be highlighted through HIDs. Figure~\ref{fig:LIVHyst1} illustrates this statement by showing the change in the gamma-ray hysteresis pattern for the flat intrinsic trend when additional LIV effects are injected with various values of $\tau$. At low fluxes, the hardness is greatly affected leading to dramatically different values, and slowly converges as the flux increases where hysteresis patterns then almost overlap. The same qualitative results hold for the two other increasing and decreasing trends (not shown here). Figure~\ref{fig:LIVHyst2} shows the HID for the same flare in the X-ray and gamma-ray ranges, with additional LIV effects of $\tau = + 400$~s\,TeV$^{-1}$ and $\tau = -400$~s\,TeV$^{-1}$, and offers a better view of the resolved pattern at high fluxes. The hysteresis pattern
for X-ray energies (blue) is unaffected by LIV, as expected , 
and can be used as reference for a flat intrinsic trend at high flux. Conversely, the gamma-ray hysteresis pattern can be strongly modified (here in particular for the superluminal case with $\tau < 0$) and in both cases shows at high fluxes a clear change in trend of the total time delays due to LIV effects, with the following behaviour: 
\begin{itemize}
\item subluminal LIV effects ($\tau>0$) induce an anticlockwise orientation of the hysteresis loop, and increasing time delays;
\item superluminal LIV effects ($\tau<0$) induce a clockwise orientation of the hysteresis loop, and decreasing time delays. 
\end{itemize}
This illustrates how $\tau$ can deeply modify the hysteresis pattern and even reverse orientation of the hysteresis loop at high fluxes when its amplitude is large enough. As a qualitative consequence, the detection of hysteresis loops with opposite orientations at high fluxes in the X-ray and gamma-ray domains would strongly hint at a contribution other than intrinsic leptonic effects being at play, and should be further investigated as a possible LIV signature.

\section{Conclusion and perspectives}\label{sec:conc}

This article was focused on energy-dependent time delays, arising as one of the most explored signatures of Lorentz invariance violation in astrophysics in the electromagnetic sector. Intrinsic time delays generated in situ by radiative processes of cosmic sources can interfere with these studies and need to be properly disentangled from propagation effects. It would indeed be necessary to accurately discriminate and quantify each contribution to any detected time delay to provide a legitimate constraint on the energy scale $E_{QG}$, and thus a proper interpretation on the tested quantum gravity models.

We investigated this issue with a time dependent modelling of blazar flares induced by particle acceleration and relying on standard leptonic one-zone scenarios. Such models naturally generate intrinsic energy-dependent time delays whether considering SSC models, with or without adiabatic expansion or compression effects, or additional external inverse-Compton emission. The core of the present work focused on a multi-wavelength study that highlights a strong correlation and symmetry between intrinsic time delays in the X-ray (unaffected by LIV) and gamma-ray (where LIV effects could arise) energy domains. This correlation is due to the fact that in leptonic scenarios the same lepton population is at the origin of the two radiative processes. Two complementary approaches were then proposed to explore and quantify that correlation especially for the pure SSC case where it is the strongest, based on two observables, namely $(i)$ the Euclidean distance between the energy-dependent time delays at X-ray and very high energies, and $(ii)$ the hardness-intensity diagrams built from spectral energy distributions that show hysteresis patterns in the two domains. From these two methods, the presence of any independent external phenomenon modifying time delays such as LIV should be revealed by the weakening or even breaking of the expected intrinsic X-gamma correlation. Moreover, LIV should induce a systematic effect always affecting the correlation in the same direction, with an amplitude varying as a function of the redshift of the cosmic sources.

The present study illustrates how all basic kinds of one-zone leptonic flares (SSC, with or without EIC and adiabatic effects) induced by particle acceleration can grow in two main regimes at high energies, either an acceleration-driven regime or a cooling-driven regime, depending on the high-energy lepton population evolution.
The evolution of intrinsic time delays as a function of the energy follow three specific trends, described here as increasing, flat, and decreasing, which can be explained from the competition between acceleration and radiative processes. The flat trend corresponds to the thin transition between the two main regimes and is obtained only with a fine-tuning of the model parameters. EIC and adiabatic processes do not modify this general picture. However, their presence influences the nature of the flare compared to a pure SSC scenario, and can even change its regime. Additional EIC radiation increases radiative losses and always pushes the flares towards the decreasing trend and cooling-driven regime at VHE. Conversely, adiabatic expansion slows down particle acceleration processes and pushes the flares towards the increasing trend and the acceleration-driven regime (and the opposite for adiabatic compression).
For these reasons, and for sake of simplicity, we mostly applied here the Euclidean distance and HID techniques to pure SSC flares. LIV effects acting only at VHE can reduce and even delete in some cases the strong correlation expected between the X-ray and VHE gamma-ray emission in leptonic scenarios. For a large domain of flare parameters, a minimum normalised Euclidean distance between delays in X-rays and in gamma-rays ($d_{E,\mathrm{min}} < 0.6$) was deduced for SSC models of sources at low redshift (z = 0.03). Therefore, obtaining values of $d_{E,\mathrm{min}} > 0.6$ from simultaneous X-ray and VHE observations of SSC flares from such sources would sign the presence of additional delays, possibly due to LIV effects. Similarly, the injection of LIV effects can significantly modify the HID hysteresis patterns at VHE energies, and can even invert the direction of rotation of the gamma-ray hysteresis curve compared to the X-ray curve. The simple detection of opposite directions of rotation of the hysteresis curve at high fluxes in X-rays and gamma-rays in SSC flares would indicate the presence of non-intrinsic time delays, possibly due to LIV.

Euclidean distance and HID approaches may be more or less accurate and conclusive to discriminate between the origins of the delays, depending on the observed events and on the instrumental performance of the present or future detectors in terms of sensitivity, and spectral and temporal resolution. They both emphasise the decisive importance of simultaneous X-ray and VHE gamma-ray monitoring of variable blazars and of efficient global multi-wavelength alarm networks to catch both rise and decay phases of strong flares. Only under this condition will we succeed in disentangling possible LIV time delays from intrinsic ones with individual sources.

The two techniques proposed here are closely related but complementary and can be adapted depending on the characteristics of available datasets. In particular, a high activity episode sometimes shows multiple flux peaks, such as in the case of the flare of blazar \object{PKS 2155-304} in 2006 \citep{Aharonian2007}. Under these conditions, it is reasonable to assume that the multiple peaks will have similar intrinsic properties. The $d_E$ technique is based on the analysis of X-ray and gamma-ray light curves in several spectral bands. It should be possible to apply it to a single peak or to multiple peaks, as long as it is possible to measure the lag as a function of energy in the X-ray and VHE domains with a high enough accuracy. On the other hand, the HID method directly reflects the evolution of the SEDs and relies on the accurate measurement of the flux and spectral index variation in time. This accuracy will probably be achieved only by analysing individual peaks separately, provided the photon statistics are high enough. A first feasibility study for the use of HIDs with CTA is currently ongoing \citep{Rosales2024b}.

Datasets on blazar flares presently available with simultaneous X-ray and VHE observations with good time coverage are still very rare, and significant time delays or accurate hysteresis patterns at VHE are still difficult to obtain, even for exceptional events \citep[see e.g.][]{Abramowski2012}. However, observations at lower energies, particularly in X-rays, suggest that the majority of blazar flares tend to be in the cooling-driven regime with a decreasing trend of intrinsic delays at high fluxes \citep{Hyst_obs1,Model4,Hyst_obs2,Hyst_obs3,Hyst_mod,Abe2023}, while a few cases are also observed in the acceleration-driven regime with an intrinsic delay increasing trend \citep{Ravasio2004,Aharonian2009}. Since observational constraints are very stringent on superluminal LIV effects \citep[see e.g.][]{lhaaso2021, COST}, it is fair to restrict the present discussion to subluminal effects. The latter would be opposite to the intrinsic trend of the cooling-driven family of flares and would tend to suppress the time delays observed at VHE or send them towards the increasing trend. In this case, an analysis of simultaneous data in X-rays and gamma-rays could lead to the detection of a clockwise (respectively anticlockwise) orientation of the hysteresis pattern in X-rays (respectively VHE gamma-rays) and the measurement of large Euclidean distances, which would provide prime clues for a LIV signal.

It is important to emphasise that intrinsic delays as well as Euclidean distance and HIDs strongly constrain leptonic models, in addition to what has to date mostly been done only with light curves and SEDs. Using leptonic models including parametrised LIV effects, it could also be possible, in principle, to remove degeneracy between LIV and intrinsic delays by identifying the scenario which fits best the whole set of observable quantities. At first glance, intrinsic one-zone SSC flares, assuming that they can be firmly identified as such by multi-wavelength and multi-messenger observations, will be the most appropriate for LIV versus intrinsic delay disentangling, due to their limited number of free parameters. The correlation between X-rays and VHE gamma-rays often observed during BL Lac flaring states has been a long standing argument in favour of SSC emission \citep[see e.g.][]{Sol2013}, and such a one-zone SSC framework still appears to provide a reasonable first-order description, in the hard X-ray and VHE ranges, of some of the brightest and best observed flares detected in archetypal high-frequency peaked BL Lac objects, such as \object{PKS 2155-304}, \object{Mkn 501}, and \object{Mkn 421} \citep{Abramowski2012,Hovatta2015,Ahnen2018,Dmytriiev2021,Abe2023}. Although they are rare events, the development of efficient alert networks and real-time analysis, and the use of data-driven detection techniques (e.g. deep learning, and anomaly detection) should promote better anticipation of the onset of flares and make it possible to monitor and characterise more of them.

However, the constantly increasing quality and complexity of the multi-wavelength datasets gathered by new generation instruments constrain the flare scenarios more and more and it will be necessary in the future to go beyond the simplistic one-zone SSC paradigm. 
In particular, the investigation of additional intrinsic effects in multi-zone scenarios, which would require describing and constraining light travel effects, and in hadronic or lepto-hadronic models \citep{Sol2022} will be a crucial development. The evaluation of Euclidean distance and HID methods in these more complex scenarios will still be needed. However, multi-zone scenarios and hadronic processes might face difficulties in reproducing fast-varying flares. Beyond these possibilities, alternative interesting proposals suggest that some flares could be due to pure geometrical phenomena inducing sudden changes in the Doppler factor of the emitting zone. Such models were developed in the literature for a few blazars showing periodicities or helical jet structures \citep{Mohan2015,Raiteri2017,Sobacchi2017,Sarkar2020,Roy2022,Abe2023,ChenJ2024}. This type of flares would certainly deserve specific studies in the context of LIV searches, since they could provide an initial signal with possibly limited intrinsic features, or at least different from standard SSC flares.

Future VHE and X-ray instruments will eventually shed new light on this crucial quest. In particular, the development of multi-wavelength and multi-messenger astronomy and the advent of the Cherenkov Telescope Array (CTA) at VHE will provide for the first time a significant sample of blazar flares observed with good spectral and temporal resolutions and coverage, at different redshifts. This should firmly constrain the true nature of blazar flares, allow discriminating intrinsic energy-dependent time delays from LIV-induced ones, and usher in a new era for both LIV and flare modelling research.

\begin{acknowledgements}
The authors would like to thank C. Boisson, J.~M. Carmona, S.~Caroff, C.~Perennes and A. ~Rosales de Leon for insightful discussions at different stages of the work presented in this article. They also thank the anonymous referee for their useful comments.
\end{acknowledgements}






\end{document}